\documentclass[final,5p,times,twocolumn]{elsarticle}
\usepackage{verbatim}
\usepackage{listings}
\usepackage{multirow}
\usepackage{graphicx} 
\usepackage{lineno}
\linenumbers

\begin{document}
\newcommand{\diana}{{\bf Diana }}
\newcommand{\apollo}{{\bf Apollo }}
\newcommand{\Te}{\ensuremath{\rm^{130}Te} }
\newcommand{\Sn}{\ensuremath{\rm^{120}Sn} }
\newcommand{\TeO}{TeO$_2$ }
\newcommand{\onubb}{$0\nu\beta\beta$}
\newcommand{\onubbplus}{$0\nu\beta^+\beta^+$ }
\newcommand{\onubplusEC}{$0\nu\beta^+$/EC }
\newcommand{\onuECEC}{$0\nu$EC/EC }
\newcommand{\kgy}{kg$\cdot$y }
\newcommand{\bbminus}{$\beta^-\beta^-$ }
\newcommand{\bbplus}{$\beta^+\beta^+$ }
\newcommand{\bplusEC}{$\beta^+$/EC }
\newcommand{\ECEC}{EC/EC }
\newcommand{\Qvalue}{1714.8 $\pm$ 1.3 keV }
\newenvironment{code}{\vspace{5mm}\footnotesize\verbatim}{\endverbatim\normalsize\vspace{5mm}} 
%
%

\title{CUORE crystal validation runs: results on radioactive contamination and extrapolation to CUORE background}
\author[INFNMilano]{F.~Alessandria}
\author[Como,INFNMiB]{E.~Andreotti\fnref{fn1}}
\author[MilanoPoli]{R.~Ardito}
\author[Milano]{C.~Arnaboldi}
\author[USC]{F.~T.~Avignone~III}
\author[LNGS]{M.~Balata}
\author[USC]{I.~Bandac}
\author[BerkeleyPhys,LBNLNucSci,LNGS]{T.~I.~Banks}
\author[INFNBologna]{G.~Bari}
\author[LBNLMatSci]{J.~W.~Beeman}
% ...

\address[INFNMilano]{INFN - Sezione di Milano, Milano I-20133 - Italy}
\address[Como]{Dipartimento di Fisica e Matematica, Universit\`{a} dell'Insubria, Como I-22100 - Italy}
\address[INFNMiB]{INFN - Sezione di Milano Bicocca, Milano I-20126 - Italy}
\address[MilanoPoli]{Dipartimento di Ingegneria Strutturale, Politecnico di Milano, Milano I-20133 - Italy}
\address[Milano]{Dipartimento di Fisica, Universit\`{a} di Milano-Bicocca, Milano I-20126 - Italy}
\address[USC]{Department of Physics  and Astronomy, University of South Carolina, Columbia, SC 29208 - USA}
\address[LNGS]{INFN - Laboratori Nazionali del Gran Sasso, Assergi (L'Aquila) I-67010 - Italy}
\address[BerkeleyPhys]{Department of Physics, University of California, Berkeley, CA 94720 - USA}
\address[LBNLNucSci]{Nuclear Science Division, Lawrence Berkeley National Laboratory, Berkeley, CA 94720 - USA}
\address[INFNBologna]{INFN - Sezione di Bologna, Bologna I-40127 - Italy}
\address[LBNLMatSci]{Materials Science Division, Lawrence Berkeley National Laboratory, Berkeley, CA 94720 - USA}
% ...

\fntext[fn1]{Presently at: Joint Research Center, Institute for Reference Materials and Measurements, 2440 Geel - Belgium}
\fntext[fn2]{Presently at: Nikhef, 1098 XG Amsterdam - The Netherlands}
\fntext[fn3]{Presently at: Institut d'Astrophysique Spatial, 91045 Orsay - France}
\fntext[fn4]{Presently at: CEA / Saclay, 91191 Gif-sur-Yvette - France}
\fntext[fn5]{Deceased}
\begin{abstract}
The CUORE Crystal Validation Runs (CCVRs) have been carried out since the end of 2008 at the Gran Sasso National Laboratories, in order to test the performances and the radiopurity of the TeO$_2$ crystals produced at SICCAS  (Shanghai Institute of Ceramics, Chinese Academy of Sciences) for the CUORE experiment. In this work the results of the first 5 validation runs are presented. Results have been obtained for bulk contaminations and surface contaminations from several nuclides. An extrapolation to the CUORE background has been performed.
\end{abstract}

\maketitle
%


\section{Introduction}
\label{sec:intro}
The production of the CUORE~\cite{CUORE} crystals was appointed to SICCAS (Shanghai Institute of Ceramics, Chinese Academy of Sciences) and began in March 2008, with the synthesis of the \TeO powder. Given the goal of CUORE in terms of background ($<$0.01 counts/keV/kg/y at the Q-value~\cite{CUORE04}), the radiopurity of the \TeO crystals is a crucial issue.

Radioactive contaminations may come from long-lived, naturally occurring isotopes, such as $^{238}$U and $^{232}$Th and their daughters and from cosmogenic activation of the detector materials after their production. To minimize the influence of long-lived nuclei, great care is devoted to the selection of all materials and ancillaries used for the preparation of the detector. Sea level transport and underground storage of prepared crystals are necessary in order to minimize their cosmogenic activation.

A dedicated protocol~\cite{Arna10} is defined for the radio-purity related quality control of the crystal production process, 
starting from metallic tellurium synthesis to the final processing of ready-to-use \TeO crystals. Radio-purity certification procedures, involving ICP-MS (Inductively Coupled Plasma Mass Spectrometry) measurements, $\gamma$ spectroscopy with HPGe detectors and $\alpha$ spectroscopy with 
Surface Barriers Detectors (SBD), are applied in each production phase to test the above mentioned materials.

At the same time, cryogenic tests are designed to test the ready-to-use \TeO crystals upon their arrival at LNGS.
The crystal validation is performed through experimental runs, each called CCVR (CUORE Crystal Validation Run), in which 4 crystals randomly 
chosen from a batch coming from SICCAS, are mounted in an setup similar to a CUORE single module and operated at cryogenic temperatures for a 
time period of several weeks in order to test the bolometric performance and the compliance of the crystals to the contract limits in terms of radio-purity (see Table~\ref{tab:ContractLimits}).
\begin{table}[htdp]
\begin{center}
\begin{tabular}{|c|c|c|}
\hline
Isotope  & Allowed Contamination  \\
\hline
\hline
$^{238}$U & $<$ 3 $\cdot 10^{-13}$ g/g  \\
\hline
$^{232}$Th & $<$ 3 $\cdot 10^{-13}$ g/g  \\ 
\hline
$^{210}$Pb & $<$ 1 $\cdot 10^{-5}$ Bq/kg  \\
\hline
$^{210}$Po & $<$ 0.1 Bq/kg  \\
\hline
\end{tabular}
\caption{Contamination limits for the ready-to-use \TeO crystals ~\cite{Arna10}.}
\label{tab:ContractLimits}
\end{center}
\end{table}

In this work the results of the first 5 validation runs, carried out from the end of 2008 to the middle of 2010, are presented. A summary of all CCVRs detector operation is reported in Table~\ref{tab:Summary1}. The 4 crystals tested in CCVR1 (two of them were again tested in CCVR2) were sent to LNGS by plane. This was necessary in order to ensure a fast response on the radioactivity level of the crystals. All the other crystals were transported by ship.
\begin{table}[htdp] 
\centering        
\small\addtolength{\tabcolsep}{-3pt}
\begin{tabular}{|c|c|c|}  
\hline                    
Run & Duration & Livetime [d]  \\    
\hline
\hline                  
CCVR1 & Dec 20, 2008 - Mar 9, 2009  & 59.9  \\
CCVR2 & Jun 6, 2009 - Jun 30, 2009  & 19.4 \\
CCVR3 & Nov 11, 2009 - Jan 4, 2010  & 43.05 \\
CCVR4 & Mar 31, 2010 - May 17, 2010 & 25.8 \\
CCVR5 & Aug 11, 2010 -  Oct 1, 2010  & 30.3  \\
\hline                         
\end{tabular} 
\caption{Summary of CCVR  data taking period and live time.}  
\label{tab:Summary1} 
\end{table} 

Radioactivity study was performed on the high energy region of the spectrum (above 4000 keV), where the contribution of the $\alpha$ lines from uranium and thorium decay chains is expected. Given the short range of $\alpha$ particles, their signature is a clear indication of a radioactive contamination within the crystals or on their surface. Results (mainly upper limits) are obtained for both bulk contaminations (Sec.~\ref{sec:resBulk}) and surface contaminations (Sec.~\ref{sec:resSurf}). For $^{210}$Pb, for which the study of the $\alpha$ lines was not possible, the lower energy portion of the spectrum has been used (Sec.~\ref{sec:210Pb}).

An extrapolation to the CUORE background is performed in Sec.~\ref{sec:extrCuore}.
\section{Experimental setup}
\label{sec:exp}
Each CCVR setup consists of an array of four crystals, arranged in a single floor which in a first approximation represents the single CUORE module. The four $5 \times 5 \times 5$ cm$^3$ crystals are enclosed in a pair of  Oxygen Free High Conductivity (OFHC) copper frames which serve both as mechanical support and thermal bath. The two frames are connected to each other by four small columns, also made of copper. Both frames and copper are  wrapped with several layers of polyethylene. The crystals are connected to the copper frames by small Teflon supports that represent the weak thermal conductance versus the heat sink.

Each crystal is provided with a Neutron Transmutation Doped Ge thermistor (NTD), which converts the phonon signal into a detectable  voltage pulse~\cite{NTD}. Some of the CCVR crystals are provided with two NTD thermistors. In these cases both channels are analyzed and the best performing channel (from the point of view of the energy resolution) is used for the final results. 

The CCVR experimental setup is hosted in a dilution refrigerator placed in the Hall C of National Laboratory of INFN at Gran Sasso and  operated at $\sim$ 10~mK.
A complete description can be found in~\cite{Qino08} and references therein for what concerns the electronics and the DAQ and in~\cite{HallC} for what concerns the cryogenic setup and the shields.

\section{Data analysis}
\label{sec:ana}
CCVR data processing, from raw data to the final spectra, follows the procedure described in details in~\cite{Qino10}.
 
The pulse amplitude is estimated by means of an Optimum Filter (OF) technique~\cite{Gatti} that reduces the noise superimposed to the signal, maximizing the signal to noise ratio. 

The mean detector response and the noise power spectral density, needed to build the OF transfer function, are estimated from bolometric pulses and noise samples (data samples recorded randomly and without triggered events) by a proper averaging procedure.

Gain instability corrections~\cite{Qino10} are performed using the 5407.5 keV $\alpha$ line from $^{210}$Po. This element is always present in recently grown crystals but decays away with a half life  of 138.38 days.

\begin{figure}[htbp]
  \centering
  \includegraphics[width=9cm]{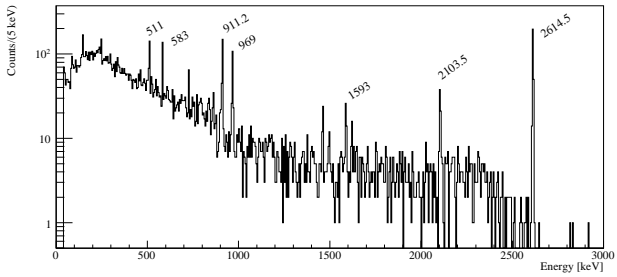}
  \caption{Calibration spectrum of Channel2 - CCVR1. Gamma lines from the $^{232}$Th decay chain are visible in the spectrum.}
  \label{fig:calib_CCVR1_ch2}
\end{figure}

The energy calibration is performed using $^{232}$Th $\gamma$ sources inserted inside the cryostat external lead shield.  
An example of calibration spectrum (Channel2 - CCVR1) is shown in Fig.~\ref{fig:calib_CCVR1_ch2}. Gamma lines from the $^{232}$Th decay chain are clearly visible in the spectrum.
The calibration is performed using a third-order polynomial function and the $^{210}$Po peak is used in addition to the  $^{232}$Th $\gamma$ peaks.

Each CCVR sum spectrum is composed by events which survive two different type of data selection: global and event-based requirements. 

Global requirements are applied following criteria decided a priori on the detector performances (an excessive noise level, ADC saturation, etc..). They identify bad time intervals that need to be discarded. These kind of cuts introduce a dead time that is accounted for by properly reducing the live time of the interested detector. 

Event-based requirements comprise: pile-up rejection, pulse-shape and coincidence selection. 

The presence of a pile-up prevents the OF algorithm from providing a correct evaluation of the pulse amplitude. The pile-up rejection is performed by imposing an extendable (paralyzable) dead window of 7 seconds to each event.  

The pulse-shape analysis is used to reject non-physical events. The pulse shape parameters are the rise time and decay time of the OF-filtered waveform and other parameters that measure the deviation of filtered raw signal from the average detector response.

\begin{figure}[htbp]
  \centering
  \includegraphics[width=8cm]{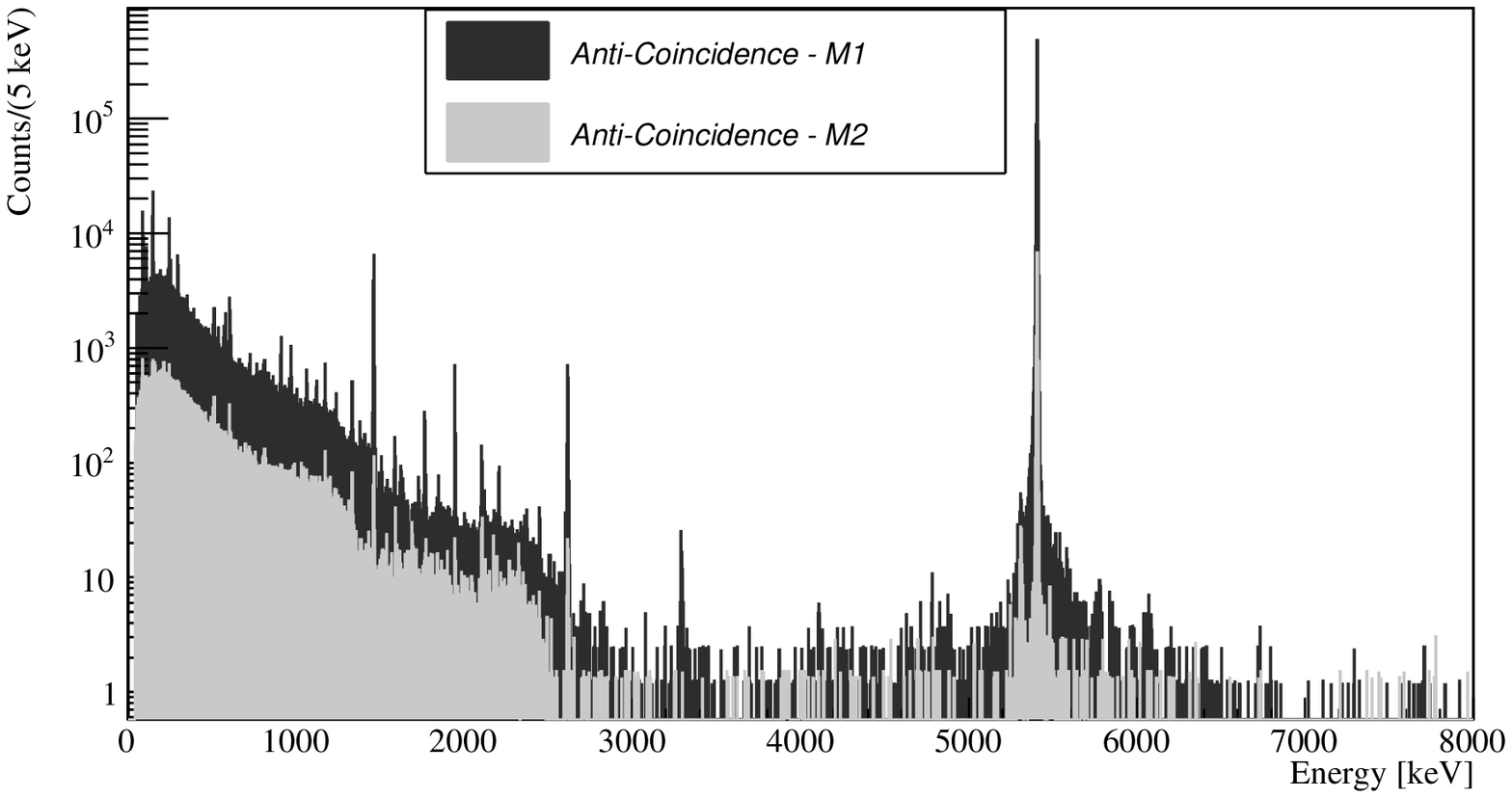}
  \includegraphics[width=8cm]{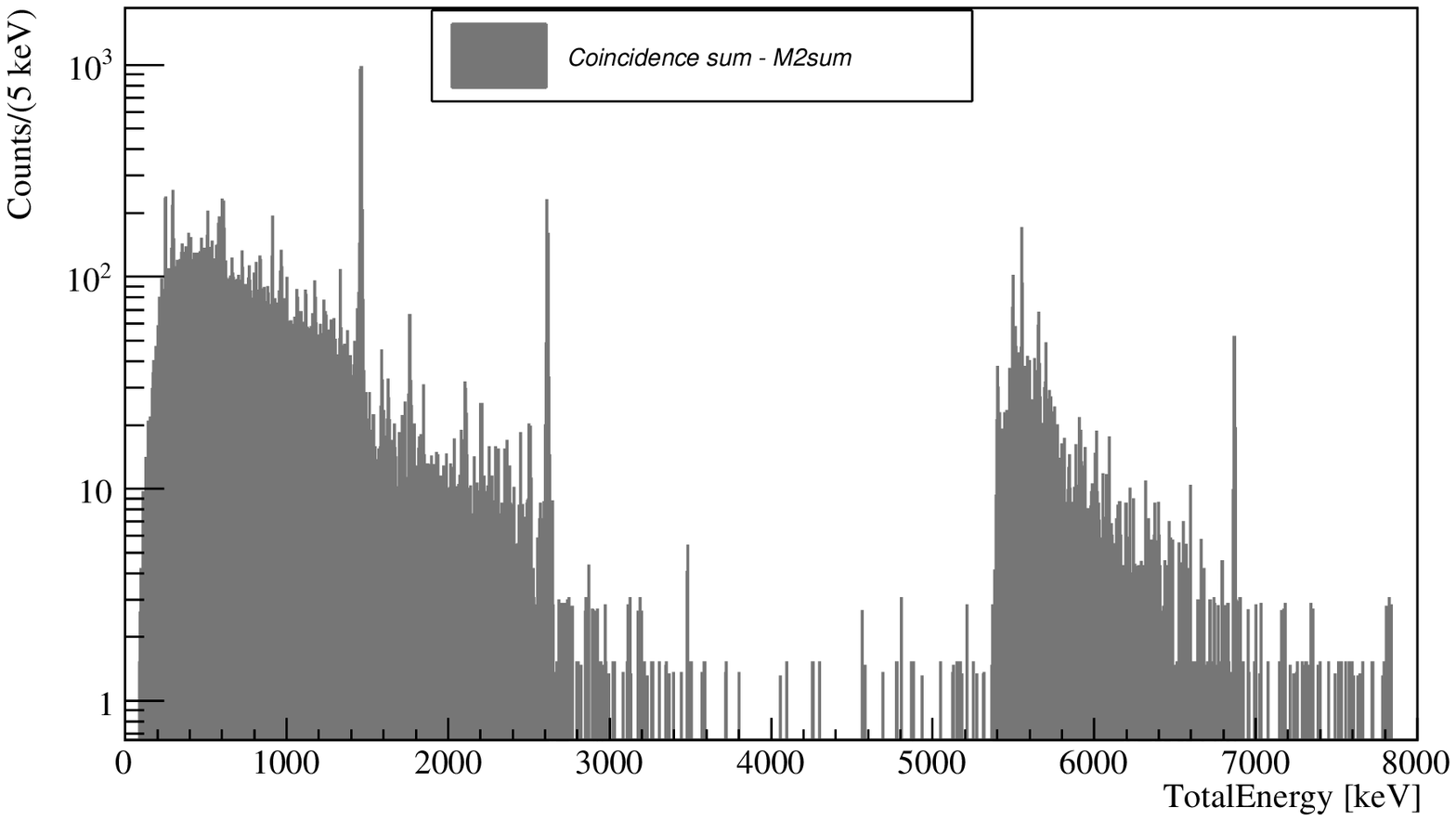}
  \caption{Energy spectra for the full CCVR statistics. Top: anti-coincidence and coincidence spectra. Bottom: Sum energy spectrum of multiplicity 2 events.}
  \label{fig:SpectrumCCVRall}
\end{figure}

As a first step, each CCVR spectrum is corrected for the corresponding efficiency of the event-based cuts (from Table~\ref{tab:Effs}) and then the spectra are summed together. Four types of spectra are produced for all CCVRs:
\begin{itemize}
\item {\bf Total Energy spectrum (TOT)}: it contains all the general cuts and the pulse shape cuts.   
\item {\bf Anti-coincidence Energy spectrum (M1)}: it contains the events that caused an energy deposition in one crystal only (anti-coincidence cut). For what concerns $\alpha$ particles, this corresponds to  bulk  events (that is decays where the emitted $\alpha$ particle  is absorbed within the crystal itself) and to  surface  events, generated by decays occurring on the surface of an inert material that hit a facing crystal, or to $\alpha$ decays occurring on a crystal surface, whose escaping products are absorbed by inert materials. 
\item {\bf Coincidence Energy spectrum (M2)}: it contains the events that caused an energy deposition in two  crystals. For what concerns $\alpha$ particles, these events arise from $\alpha$ decays occurring on a crystal surface faced to another crystal. 
\item {\bf Coincidence Sum Energy spectrum (M2sum)}: it contains the  sum energy of multiplicity 2 events. For instance in the case of an $\alpha$ decay on the surface of a crystal, both the energy of the $\alpha$ particle $E_1$ in the facing crystal and the corresponding nuclear recoil energy $E_2$ in the original crystal are detected. In the spectrum M2sum the variable $E_{TOT}=E_1+E_2$ is plotted. 
\end{itemize}
Fig.~\ref{fig:SpectrumCCVRall} shows the sum spectra of all CCVRs.

\subsection{Efficiency of event-based cuts}
Due to the high rate of $^{210}$Po events, there is a significant loss of efficiency due to pile-up rejection. The efficiency is estimated as:
\begin{equation}
\varepsilon_{\rm{pile-up}} = 1 - P_{\rm{pile-up}} = e^{( -r \cdot T )}
\label{eq1}
\end{equation}
where $P$ is the probability of a pile-up, $r$ is the counting rate of the events that passed the global cuts described above, and $T$ is the length of a time interval containing an event during which the occurrence of another event would be considered pile-up. The interval $T$ contains some time after the event during which a double pulse would result and some time before the event during which the event's baseline would be spoiled by the tail of the preceding pulse. 
Properly, $T$ depends on the energy of the other event: the higher the energy of the other event, the longer its tail remains too large.
The counting-rate is channel-dependent and sometimes also time-dependent. In the specific case of $^{210}$Po events, the counting-rate obviously decreases with time due to polonium decay (half-life: 138.38 days).

For the sake of simplicity, an average pile-up reduction efficiency for each CCVR is computed. This will apply to all channels in the full energy range and is  calculated using formula~(\ref{eq1}) with $T$ = 7 s and using as $r$ the global counting rate after the general cuts.  
\begin{table}[htdp]
\begin{center}
\begin{tabular}{|c|c|c|c|c|c|}
\hline
Run  & $\varepsilon_{\rm{pile-up}}$ & $\varepsilon_{\rm{PSA}}$ & $\varepsilon_{\rm{AC}}$ \\
\hline
\hline
CCVR1 & 0.84 $\pm$0.01 & 0.96$\pm$0.01 & 0.987$ \pm$0.003 \\ 
CCVR2 & 0.88 $\pm$0.01 & 0.98$\pm$0.01 & 0.982$\pm$0.005 \\
CCVR3 & 0.89 $\pm$0.01 & 0.97$\pm$0.01 & 0.990$\pm$0.002  \\
CCVR4 & 0.88 $\pm$0.01 & 0.94$\pm$0.02 & 0.987$\pm$0.003 \\
CCVR5 & 0.89 $\pm$0.01 & 0.98$\pm$0.01 & 0.990$\pm$0.003 \\
\hline
\end{tabular}
\caption{Efficiencies of event-based cuts for each CCVR.}
\label{tab:Effs}
\end{center}
\end{table}

The pulse shape cuts efficiency $\varepsilon_{\rm{PSA}}$ is evaluated  on the background peak at 2614.5 keV due to $^{208}$Tl, by a simultaneous fit on both the spectra of accepted and rejected events as detailed in \cite{Qino10}.

The same procedure is applied  for the evaluation of  the anti-coincidence cut efficiency $\varepsilon_{\rm{AC}}$. In this case, instead of the 2614.5 keV line (which is usually in coincidence with other $\gamma$ lines), the photopeak at 1460.8 keV due to $^{40}$K is used. Results are summarized in Table~\ref{tab:Effs}. 
\subsection{Monte Carlo simulations}
\label{sec:MCsim}
In order to extract numerical information on the activity or contamination of a given nuclide in CUORE crystals from CCVRs data, it is necessary to rely on Monte Carlo simulations, capable of reproducing the main features of the detector geometry and response.
 
CCVRs simulations are performed with the GEANT4-based code developed by the CUORICINO ad CUORE collaboration and described in~\cite{ArtRadio} and~\cite{ArtChambery}. The simulation takes into account the energy resolution and the threshold of each detector. For bulk contaminations, an homogeneous distribution of the given nuclide inside the crystals is assumed. For surface contaminations, different depths of the contamination layer (from 0.01 $\mu$m to 10 $\mu$m) are considered and the contamination density profile is assumed to decrease exponentially with the depth from the surface to the bulk of the crystal. 

In CCVR1 there is a not negligible number of measurements ($\sim$ 4\%) with only 2 active facing crystals. This affects the efficiency calculation of surface contaminations. To take into account this variation from the standard configuration with 4 crystals, 2 different efficiencies are calculated for each Monte Carlo simulation:
\begin{itemize}
\item $\varepsilon^{2ch}_{MC}$: corresponding to the CCVR1 configuration with only 2 active crystals;
\item $\varepsilon^{4ch}_{MC}$: corresponding to the standard CCVR configuration, with 4 active crystals;
\end{itemize}
The average efficiency, weighed on the proper lifetime $\Delta t^i$, is computed using the formula:
\begin{equation}  \label{eq:effMC}
  \varepsilon_{MC}= \frac{\varepsilon^{2ch}_{MC} \; \Delta t^{2ch} + \varepsilon^{4ch}_{MC}  \;  \Delta t^{4ch}} { \Delta t^{2ch} +\Delta t^{4ch}}
\end{equation}
\section{Results on background rates of CUORE crystals}
\label{sec:resRates}
From the energy spectra of all CCVRs the background rates in various energy regions can be calculated.

Six energy regions of interest are identified in the spectra and the corresponding count rates for anticoincidence (M1) and coincidence (M2) spectra are calculated. Results for the global spectra are reported in Table~\ref{tab:CCVRsRates} (errors are statistical only). 
\begin{table}[htdp]
\begin{center}
\small\addtolength{\tabcolsep}{-1pt}
\begin{tabular}{|c|c|c|c|}
\hline
        & Continuum       & $^{190}$Pt       & Continuum   \\
        & (2700, 3200)    & (3200, 3400)    & (3400, 3900) \\
        & [keV]           & [keV]           & [keV]        \\ 
\hline
M1      & 0.19$\pm$0.02   & 0.38$\pm$0.04   & 0.09$\pm$0.01 \\ 
M1-PoSub & 0.13$\pm$0.02   & 0.34$\pm$0.04   & 0.06$\pm$0.01 \\
\hline
M2      & 0.05$\pm$0.01   & 0.02$\pm$0.01   & 0.025$\pm$0.006 \\
M2-PoSub & 0.02$\pm$0.01   & 0.01$\pm$0.01 & 0.008$\pm$0.008 \\
\hline
\hline
        & U/Th            & $^{210}$Po       & U/Th        \\
        & (4000, 5000)    & (5000, 6000)    & (6000, 8000) \\
        & [keV]           & [keV]           & [keV]        \\
\hline
M1      & 0.19$\pm$0.01   & -               & 0.057$\pm$0.004 \\
M1-PoSub & 0.13$\pm$0.01   & -               & 0.057$\pm$0.004 \\
\hline
M2      & 0.04$\pm$0.01   & -               & 0.014$\pm$0.002 \\
M2-PoSub & 0.014$\pm$0.007 & -               & 0.014$\pm$0.002 \\
\hline
\end{tabular}
\caption{Count rates measured in [counts/keV/kg/y]. Errors are statistical.}
\label{tab:CCVRsRates}
\end{center}
\end{table}

The continuum region (2700, 3200) keV is of great interest since it is the region immediately above the Q-value of the neutrinoless double beta decay of \Te ~\cite{Qino10}.

In the region (3200, 3400) keV a contribution of the $\alpha$ line from $^{190}$Pt is expected. This contamination is almost unavoidable for \TeO crystals, as explained in Sec.~\ref{sec:resBulk}.

From 4000 to 8000 keV the contribution of the various $\alpha$ lines from U and Th decay chains is expected.

In between, there is the region (5000, 6000) keV, which is affected by the $^{210}$Po contamination (see Sec.~\ref{subsec:210Po}). This produces not only a peak at the energy of the Po $\alpha$ line 5407.5 keV but also a broad background over the entire region due to mis-identified pile-up events (above the peak energy) or to the escape of the $\alpha$ that releases part of its energy in a inert material (below the peak energy). An indication of the rate in this region is of no particular interest, also because of the relative short half-life of $^{210}$Po (138.38 days) that guarantees a huge reduction of this count rate when CUORE will start the data taking.

\subsection{Subtraction of $^{210}$Po induced rate}\label{sec:PoSub}
Because of the presence of $^{210}$Po, an excess of count rate could arise in the M1 and M2 spectra below the energy of the 5407.5 keV $\alpha$ line, if the contamination of $^{210}$Po is close enough to the surface for the $\alpha$ to escape and release part of its energy in an inert material (M1 spectrum) or in a nearby detector (M2 spectrum). This contribution from the M2 spectrum can be estimated, calculating for each energy region the rate of M2 events in which the total energy $E_{TOT}$ lies in the interval (5407.5$\pm$50) keV. The M2 count rate subtracted for this contribution is defined as M2-PoSub and reported in Table~\ref{tab:CCVRsRates} for comparison.
 
In a similar way the contribution of surface $^{210}$Po in the M1 spectrum is evaluated. Because of the geometry of the CCVR setup, our coincidence analysis is sensitive only to 8 faces over 24. So the anti-coincidence count rate due to $^{210}$Po  of the remaining 16 faces is assumed to be two times the coincidence one, calculated on the M2 spectrum as described above.\footnote{In some of CCVR1 measurements ($\sim$ 4\% of the CCVR lifetime) only two facing crystals were active. The extrapolation of M2 counts to M1 is performed considering that the coincidence analysis is sensitive only to 2 over 12 faces.}
The M1 count rate subtracted for the surface $^{210}$Po contribution is defined as M1-PoSub and reported in Table~\ref{tab:CCVRsRates}.

\subsection{Background rates comparison with previous \TeO detectors}
It is interesting to compare CCVRs rates with the ones obtained with previous detectors, like TTT~\cite{TTT} or Cuoricino~\cite{Qino10}. This comparison is shown in Table~\ref{tab:Comp}, where the M1 and M2 count rates for CCVRs are after polonium subtraction, as described in ~\ref{sec:PoSub}.
\begin{table}[htdp]
\begin{center}
\begin{tabular}{|c|c|c|c|}
\hline
           &     & Continuum        & U/Th         \\
           &     & (2700, 3900)     & (4000, 5000) \\
           &     &  [keV]           & [keV]        \\
\hline
CCVR      & M1  & 0.09$\pm$0.02    & 0.13$\pm$0.01  \\
           & M2  & 0.015$\pm$0.007  & 0.014$\pm$0.003 \\
\hline
TTT      & M1  & 0.052 $\pm$0.008   & 0.28$\pm$0.02  \\
           & M2  & 0.009$\pm$0.003  & 0.0018$\pm$0.005 \\
\hline
Cuoricino  & M1  & 0.104 $\pm$0.002 & 0.522$\pm$0.003 \\
           & M2  & 0.009$\pm$ 0.001 & 0.084$\pm$0.001 \\
\hline
\hline
           &     & $^{210}$Po        & U/Th         \\
           &     & (5000, 6000)     & (6000, 8000) \\
           &     & [keV]            & [keV]        \\
\hline
CCVR       & M1  & -                & 0.057$\pm$0.004 \\
           & M2  & -                & 0.014$\pm$0.002 \\
\hline
TTT      & M1  & 1.30$\pm$0.07      & 0.025 $\pm$ 0.004 \\
           & M2  & 0.09$\pm$0.01     & 0.005 $\pm$ 0.002 \\
\hline
Cuoricino  & M1  & 0.846$\pm$0.004   & 0.099 $\pm$ 0.001 \\
           & M2  & 0.173$\pm$0.002   & 0.0163 $\pm$ 0.0004 \\
\hline
\end{tabular}
\caption{Count rate comparison with previous detectors, measured in [counts/keV/kg/y]. Here CCVR values are after polonium subtraction, as described in ~\ref{sec:PoSub}. Errors are statistical.}
\label{tab:Comp}
\end{center}
\end{table}

The notation (2700, 3900) refers to the combination of both the continuum region (2700, 3200) keV and (3400, 3900) keV,  excluding the $^{190}$Pt  energy region of (3200, 3400) keV.
It can be inferred that :
\begin{itemize}
\item in the region (2700, 3900) keV the CCVRs anti-coincidence rate is compatible within 1.8 $\sigma$ with the corresponding TTT value.

\item in the region (4000, 5000) keV, as already measured in the TTT run, a reduction in the count rate with respect to Cuoricino is observed. 

\item in the region (5000, 6000) keV a comparison is not possible since CCVR is affected by the high rate of $^{210}$Po. This is due to the fact that in CCVRs only recently grown crystals are measured, unlike TTT or Cuoricino. 

\item in the region (6000, 8000) keV, as already measured in the TTT run, a reduction in the count rate with respect to Cuoricino is present. However, the CCVR count rate is greater than the TTT value, probably because of the presence of mis-identified pile-up (in M1) or coincidences (in M2) with $^{210}$Po events, extending above 6000 keV. This contribution should decay away with $^{210}$Po.
\end{itemize}

A more detailed evaluation of the comparison can be found in~\cite{TTT}.
\section{Results on bulk contaminations of CUORE crystals}
\label{sec:resBulk}
Bulk contaminations in CUORE crystals are expected from:
\begin{itemize}
\item $^{210}$Po, as a result of the chemical affinity between polonium and tellurium;  
\item natural contaminants, like $^{238}$U and $^{232}$Th with their radioactive chains;
\item $^{210}$Pb, as a result of the deposition of lead nuclei produced by $^{222}$Rn decays, during the crystal handling in free atmosphere;
\item $^{190}$Pt, due to the fact that platinum is used in several phases of the crystal production cycle. Platinum crucibles are used for the calcination of TeO$_2$ powder used for the crystal growth and the growth crucibles are made of platinum foil. The central part of the as-grown crystal ingot is selected for the CUORE crystals in order to avoid the risk of platinum contamination on the surface of the crystal due to possible diffusion during the growth process. Possible Pt contaminations are therefore in the bulk of the CUORE crystals. 
\end{itemize}

For $^{210}$Po the activity is determined from a fit to the anti-coincidence rate vs. time behaviour (see Sec.~\ref{subsec:210Po}).

For $^{238}$U and $^{232}$Th (see Sec.~\ref{subsec:UThbulk}), the limit on the level of contamination is determined from the intensities of the $\alpha$ peaks in the anti-coincidence spectrum of all crystals, or from integrals centered at the peak position if the peaks are not visible. This is because an $\alpha$ decay from bulk contamination releases the entire Q-value of the reaction ($\alpha$ energy + nuclear recoil) in a single crystal. 

For $^{210}$Pb, the limit on bulk contamination is determined from a fit in the energy region (40, 60) keV in a subset of CCVRs data with high statistics and low threshold (see Sec.~\ref{sec:210Pb}).

\subsection{$^{210}$Po bulk activity} \label{subsec:210Po}
The activity of $^{210}$Po can be measured from the intensity of 5407.5 keV $\alpha$ line in the anti-coincidence spectrum.

\begin{figure}[htbp]
  \centering
  \includegraphics[width=9cm]{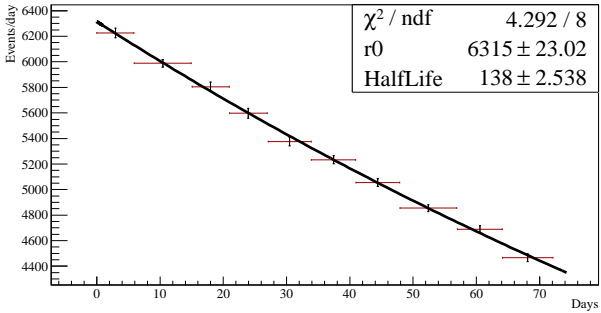}
  \caption{Global rate of $^{210}$Po events over time for each CCVR1. The fit result is overlayed.}
  \label{fig:Po210}
\end{figure}

The plot in Fig.~\ref{fig:Po210} shows the global rate of $^{210}$Po events over time for CCVR1. For each channel the $^{210}$Po events are selected in a $\pm$ 20 keV window around the energy of the $\alpha$ line. 

The units on the x-axis are days since the start of the first background measurement. Each point represents a group of measurements
whose livetime is at least 5 days.

The horizontal error bars indicate the beginning and the end of each group of runs, they are for visualization only.

Each point is corrected with the corresponding rate-based efficiency, calculated with the equation~(\ref{eq1}). A larger dead time window (9 seconds instead of 7 seconds) compared to the rest of the analysis is used, to be more conservative in the removal of pile-up pulses.

The fit function is a pure exponential:
\begin{equation}
r(t) = r_0 e^{( -ln2\;t/T_{1/2} )}
\label{eq3}
\end{equation}
where $r_0$ is the rate at the beginning of the measurement and $T_{1/2}$ is the $^{210}$Po half-life.
The half-life of the exponential decay has been evaluated for all CCVRs and it is shown in Tab~\ref{tab:PoHalfLife}: it is in good agreement (1 $\sigma$) with the half-life of $^{210}$Po (138.38 d). This indicates that the $^{210}$Po contamination is out of equilibrium and it is not being fed by $^{210}$Pb.

\begin{table}[htdp]
\begin{center}
\begin{tabular}{|c|c|}
\hline
Run & Half-life    \\
    & [days] \\
\hline
\hline
CCVR1   &  138.0$\pm$ 2.5 \\
CCVR2  & 132.3 $\pm$ 14.7 \\ 
CCVR3   & 139.4 $\pm$ 6.3 \\
CCVR4   & 137.8 $\pm$ 11.5 \\
CCVR5   & 136.6 $\pm$ 9.1 \\
\hline
\end{tabular}
\caption{$^{210}$Po half-life for all CCVRs.  The activity has been fitted using an exponential function. All the values are consistent (within 1 $\sigma$) with 138.38 days.}

\label{tab:PoHalfLife}
\end{center}
\end{table}

From the value of $r_0$ returned by the fit for each CCVR and for each crystal, the $^{210}$Po activity at the beginning of the measurement is extracted as follows:
\begin{equation}
A {\rm [Bq/kg]} = \frac{r_0}{86400{\rm [s/day]} \; m {\rm [kg]}}
\label{eq4}
\end{equation}
where $m$ is the crystal mass.

Knowing the time elapsed since the "crystal birth date" (growth completed, before the cut and shape) and the start of the measurement, the $^{210}$Po activity at production is computed.

The $^{210}$Po activities for each CCVR and for all crystals are reported in Table~\ref{tab:Po210}. The results for crystals 007 and 011, measured first in CCVR1 and again in CCVR2, are consistent.  

All crystals tested in the first five CCVRs are well below the limit of 0.1 Bq/kg imposed to the crystal producers.
\begin{table}[htdp]
\begin{center}
\begin{tabular}{|c|c|c|}
\hline
CCVR & Crystal & $^{210}$Po activity [Bq/kg]   \\
\hline
\hline
1  & 041 & 0.0257 $\pm$ 0.0001 \\
1  & 011 & 0.0510 $\pm$ 0.0005 \\ 
1  & 039 & 0.0229 $\pm$ 0.0001 \\
1  & 007 & 0.0414 $\pm$ 0.0004 \\
\hline
\hline
2  & 076 & 0.021 $\pm$ 0.004 \\
2  & 011 & 0.07 $\pm$ 0.02 \\
2  & 096 & 0.055 $\pm$ 0.006 \\
2  & 007 & 0.047 $\pm$ 0.015 \\
\hline
\hline
3  & 190 & 0.0078 $\pm$ 0.0005 \\
3  & 236 & 0.0159 $\pm$ 0.0005 \\
3  & 180 & 0.0203 $\pm$ 0.0008 \\
3  & 229 & 0.0283 $\pm$ 0.0008 \\
\hline
\hline
4  & 340 & 0.032 $\pm$ 0.004 \\
4  & 313 & 0.005 $\pm$ 0.001 \\
4  & 354 & 0.039 $\pm$ 0.004 \\
4  & 380 & 0.040 $\pm$ 0.004 \\
\hline
\hline
5  & 455 & 0.019 $\pm$ 0.002 \\
5  & 416 & 0.024 $\pm$ 0.003\\
5  & 436 & 0.032 $\pm$ 0.003 \\
5  & 421 & 0.020 $\pm$ 0.003 \\
\hline
\end{tabular}
\caption{$^{210}$Po activity at production time for all CCVRs crystals.}
\label{tab:Po210}
\end{center}
\end{table}
\subsection{U/Th bulk contaminations}
\label{subsec:UThbulk}
$^{238}$U and $^{232}$Th bulk contaminations are evaluated from the anti-coincidence sum spectrum of all CCVRs.
For each of the peaks reported in Table~\ref{tab:UThbulkNoC}, the number of counts (corrected by the efficiency of the event-based cuts from Table~\ref{tab:Effs}) within an energy window of $\pm$6 $\sigma$ around the $\alpha$ Q-value is estimated. A $\sigma$ of 2.2 keV, corresponding to the average between the values of the $^{210}$Po peak $\sigma$s for each CCVR, weighted by the corresponding lifetime of that run (see Table~\ref{tab:sigmaPo} for details) is used. 
Using the Bayesian approach, the upper limits $N_u$ at 90\% C.L. are obtained,  assuming 0 expected background counts and a flat prior for the signal ~\cite{PDG} (see Table~\ref{tab:UThbulkNoC}).

\begin{table}[htdp]
\begin{center}
\begin{tabular}{|c|c|c|c|c|}
\hline
Chain     & Nuclide   & Energy & Half-life &  $N_u$ \\
          &           & [keV]  & & 90\% C.L.     \\
\hline
\hline
$^{238}$U  & $^{238}$U  & 4270.0 & 4.47E+09 y & 11.7 \\
          	  & $^{234}$U  & 4858.8 & 2.45E+05 y & 20.8\\
          	  & $^{230}$Th & 4770.0 & 7.54E+04 y & 24.9\\
            	  & $^{226}$Ra & 4870.6 & 1599 y & 30.0\\
           	  & $^{218}$Po & 6114.7 & 3.05 min& 7.4\\
\hline
$^{232}$Th & $^{232}$Th & 4082.8 & 1.4E+10 y & 5.9\\
          & $^{212}$Bi & 6207.1 & 60.55 min & 13.5\\
    \hline
  \end{tabular}
    \caption{90\% C.L. limits on the number of events ascribed to several nuclides from uranium and thorium decay chain. For each nuclide is also shown the Q-value and the half-life of the $\alpha$ decay. }\label{tab:UThbulkNoC}
\end{center}
\end{table}

\begin{table}[htdp] 
\centering        
\begin{tabular}{|c|c|c|}  
\hline
Run   & Livetime & $\sigma$\\
      & [days]   & [keV] \\
\hline    
\hline              
CCVR1 &  59.9    &  1.6   \\ 
CCVR2 &  19.4    &   1.4  \\ 
CCVR3 &  43.05   &   2.2 \\ 
CCVR4 &  25.8    &   3.5  \\ 
CCVR5 &  30.3    &    2.8\\ 
\hline                         
\end{tabular} 
\caption{Summary of CCVRs energy resolutions, evaluated with a gaussian fit of the 5407.5 keV peak from $^{210}$Po. The average value of 2.2 keV, used for the global analysis is  obtained averaging the resolution of each CCVR with the corresponding livetime.}  
\label{tab:sigmaPo} 
\end{table} 

The upper limit on the activity for each nuclide is calculated using the following formula:
\begin{equation}  
A_u{\rm [Bq/kg]}= \frac{N_u}{\varepsilon_{MC} \; T[{\rm s}] \; m[{\rm kg}] \; \Gamma} 
\label{eq:Activity}
\end{equation}

where $\varepsilon_{MC}$ is the Monte Carlo detection efficiency, $T$ the CCVRs livetime, $m$  the crystal mass and $ \Gamma$ the branching ratio of the nuclide. In this analysis a total containment of the anti-coincidence events in the crystals is assumed ($\varepsilon_{MC}$ =1). Results are shown in Table~\ref{tab:UThbulkEq}.

The upper limit for confidence level of 90\% for U/Th bulk contaminations are then calculated in the hypothesis of secular equilibrium within the uranium and thorium decay chains. 
Results for the total CCVRs data are shown in the last column of Table~\ref{tab:UThbulkEq}. 
\begin{table}[htdp]
\begin{center}
  \begin{tabular}{|c|c|c|c|}
    \hline
    Chain & Nuclide  &  Upper limit  & Upper limit \\
          &          &    [Bq/kg] &     [g/g]   \\
    \hline
    \hline
    $^{238}$U & $^{238}$U  & 2.5E-07 & 2.0E-14  \\
                        & $^{234}$U  & 4.7E-07 & 3.6E-14   \\
                        & $^{230}$Th &  5.7E-07 &4.4E-14 \\
                        & $^{226}$Ra & 6.7E-07& 5.3E-14  \\
                        & $^{218}$Po & 1.6E-07&1.3E-14 \\
    \hline
    $^{232}$Th & $^{232}$Th & 1.3E-07& 3.1E-14\\
                         & $^{212}$Bi   & 8.4E-07 & 2.1E-13\\ 
    \hline
  \end{tabular}
  \caption{Upper limits at 90\% C.L. on the activity and on the bulk contamination of uranium and thorium decay chains in the hypothesis of secular equilibrium.}\label{tab:UThbulkEq}
\end{center}
\end{table}

In the most conservative approach, the bulk contamination limit on $^{238}$U  and $^{232}$Th is set considering the most active nuclide for each chain.
The upper limit at 90\% C.L. for uranium and thorium bulk contamination are:
\begin{eqnarray*} 
 ^{238}{\rm U} & < 5.3\cdot 10^{-14} {\rm [g/g]} \\
 ^{232}{\rm Th}& < 2.1 \cdot 10^{-13} {\rm [g/g]} 
\end{eqnarray*}
Both values of the upper limits are within the contract specification of 3$\cdot$10$^{-13}$[g/g]. For the $^{238}$U decay chain, the contribution from $^{210}$Pb is treated separately (see Sec.~\ref{sec:210Pb}).

It is important to stress the fact that the above upper limits are calculated under the hypothesis that the observed counts for each nuclide are entirely due to a bulk contamination of that nuclide. This is a conservative hypothesis since there is  not  a clear indication (for example a line) that such a bulk contamination actually exists and that the observed counts are not due to background of some other origin.

\section{$^{210}$Pb activity}
\label{sec:210Pb}
During the production of CUORE crystals great care is devoted in order to minimize the exposure of the crystals to free atmosphere, to avoid contamination of radon and its daughters. Nevertheless, a small $^{210}$Pb contamination (bulk or surface) might occur. This contamination can be very dangerous because, even if the soft beta (16.96 keV and 63.5 keV) and $\gamma$ (46.5 keV) radiation from $^{210}$Pb is generally self absorbed, the induced bremsstrahlung of the daughter $^{210}$Bi can give rise to a continuum up to 1.16 MeV. Moreover, the $\alpha$ decay of the daughter $^{210}$Po can contribute to the continuum background in the double beta decay energy region.

In CCVRs crystals the $^{210}$Pb contamination cannot be estimated from $^{210}$Po, because this contamination is out of equilibrium (see Sec.~\ref{subsec:210Po}). The only available signature is a combination of a beta spectrum (end point 16.96 keV) with a de-excitation energy of 46.5 keV (most of the cases through a conversion electron).
This produces a broad signature in the energy region (40, 60) keV, whose shape depends on the location of the Pb contamination.

\begin{figure}
   \centering		
   \includegraphics[width=8cm,height=4cm]{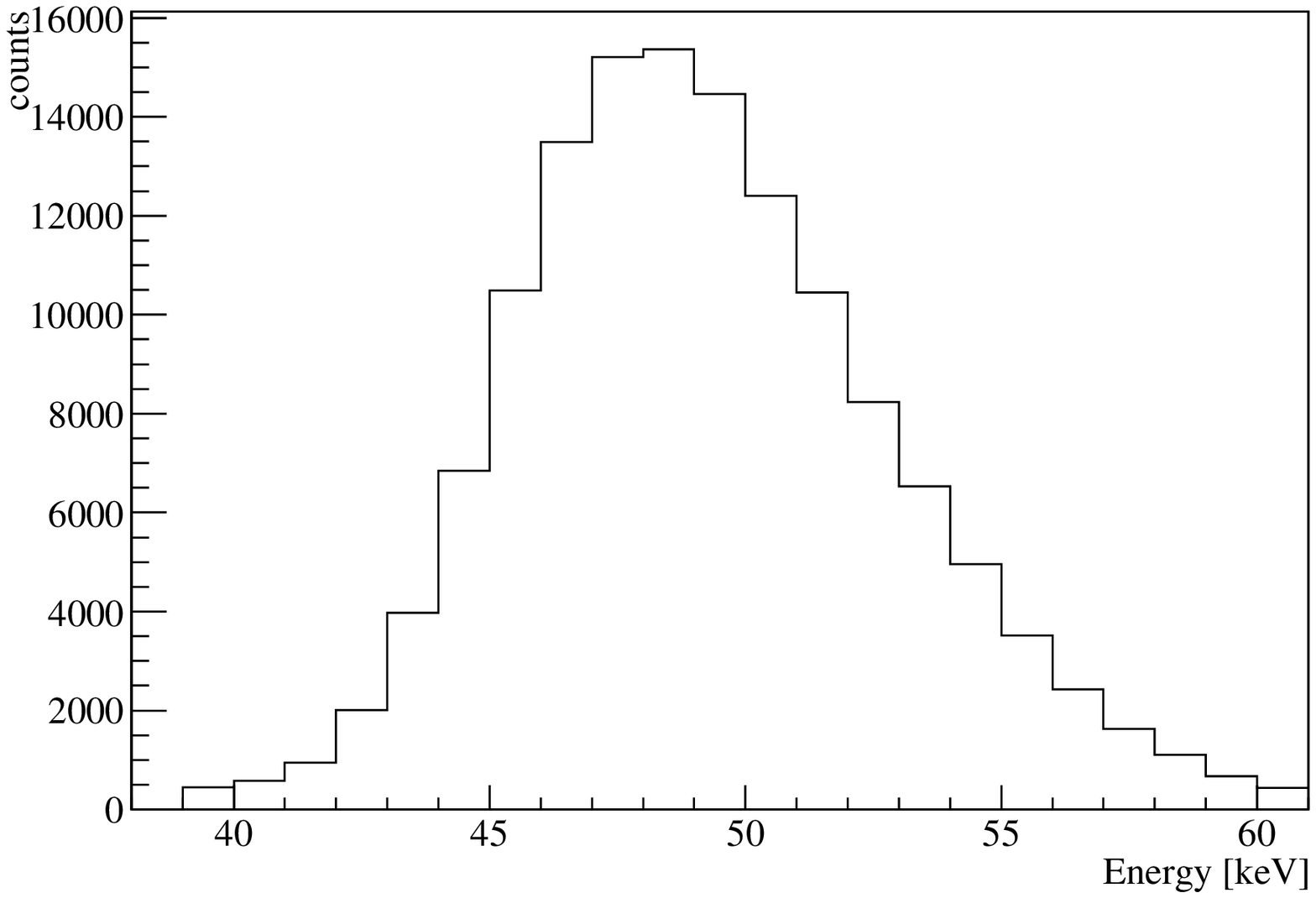}
   \includegraphics[width=8cm,height=4cm]{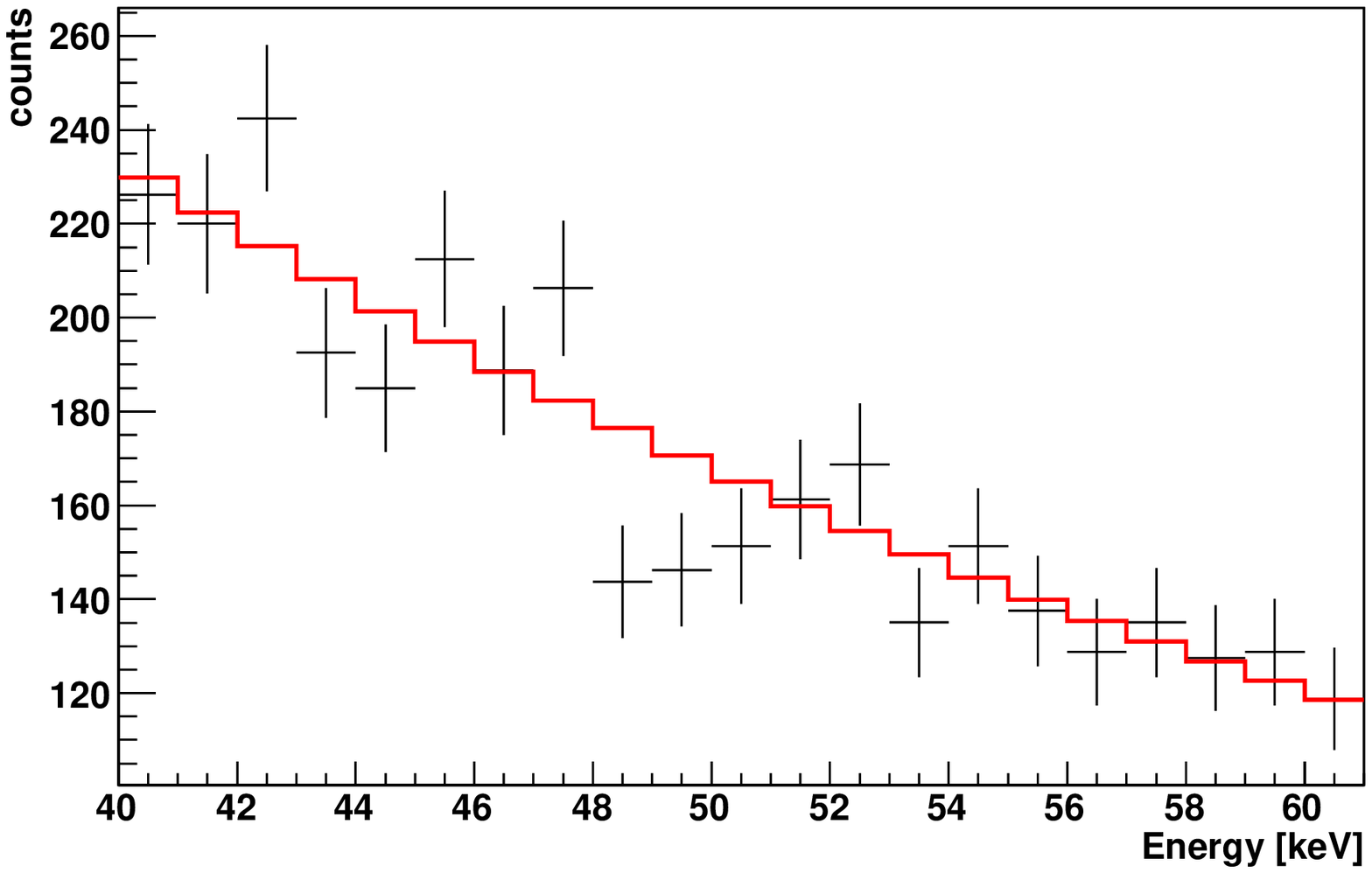}
   \caption{Top: Monte Carlo simulation of a $^{210}$Pb bulk contamination in CCVR1. Bottom: CCVR1 low energy anti-coincidence spectrum. The energy threshold is set to 40 keV. The red line represents the fit to the spectrum with an exponential background and allowing the presence of a $^{210}$Pb bulk contamination with the above shape. No hint of such contamination is found.} 
  \label{fig:CCVR1_Pb210}
 \end{figure}

Several simulations are  performed both for a bulk contamination and for a surface contamination with exponential density profile~\cite{ArtRadio}  and contamination depth varying from 0.01 $\mu$m to 10 $\mu$m. Even if, for the sake of simplicity, a common energy threshold of 50 keV is set for all CCVRs, there are subset of data where the threshold can be set to a lower value in order to look for this signature. 
CCVR1 data are used because they struk a balance between high livetime (CCVR1 has the highest statistic) and good bolometer performances. Only for CCVR1 data, the analysis is repeated with an energy threshold of 40 keV. The anti-coincidence spectrum between 40 and 60 keV is fitted with an exponentially decreasing background and allowing the presence of a $^{210}$Pb spectrum with shape taken from Monte Carlo simulation. The free parameters of the fit are the two parameters of the exponential background and the total number of counts from $^{210}$Pb. The fit is repeated for each Monte Carlo signature. The number of $^{210}$Pb events is compatible with zero within the error for all signatures. An example of Monte Carlo spectrum for a $^{210}$Pb bulk contamination and the corresponding fit to the experimental spectrum is shown in Fig.~\ref{fig:CCVR1_Pb210}. 

The upper limit at 90\% C.L. on the number of $^{210}$Pb counts is  $N_u$ = 1.644 $\sigma$, where $\sigma$ is the error on the number of counts returned from the fit. The upper limit on bulk contamination is computed using Eq.~\ref{eq:Activity}. The upper limit on surface contamination is extracted using the following formula:
\begin{equation}
A_u{\rm [Bq/cm^2]}= \frac{N_u}{\varepsilon_{MC} \; T[{\rm s}] \; S[{\rm cm}^2]} 
\label{eq:SurfActivity}
\end{equation}
 where $\varepsilon_{MC}$ is the Monte Carlo detection efficiency,  $T$ is the livetime and S is the surface of a crystal (150 cm$^2$). 
Results are shown in Table~\ref{tab:CCVR1Pb210}. The upper limit on the activity for bulk contamination is below the contract limit of 10$^{-5}$ Bq/kg (see Table~\ref{tab:ContractLimits}).
\begin{table}[htdp]
\begin{center}
  \begin{tabular}{|c|c|}
    \hline
   contamination  & Upper limit 90\% C.L. \\
    \hline
    \hline 
	bulk             & 3.3E-06 \\
	surf. 0.01$\mu$m & 9.8E-07 \\
	surf.  0.1$\mu$m & 3.8E-08 \\
	surf.  0.2$\mu$m & 2.2E-08 \\
	surf.    1$\mu$m & 9.2E-09 \\
	surf.    5$\mu$m & 5.6E-09 \\
	surf    10$\mu$m & 4.9E-09 \\
    \hline
  \end{tabular}
  \caption{Upper limits on the activity of bulk and surface contamination of $^{210}$Pb. Bulk contamination is given in [Bq/kg], surface contaminations are given in [Bq/cm$^2$].} \label{tab:CCVR1Pb210}
\end{center}
\end{table} 
\section{Results on surface contaminations of CUORE crystals}
\label{sec:resSurf}
Surface contaminations of CUORE crystals are expected from the same nuclides listed in Sec.~\ref{sec:resBulk}, except for $^{190}$Pt, which is all in the bulk. Uranium and thorium surface contaminations can be investigated by means of coincident events in two facing crystals: surface contamination of an $\alpha$-decaying nuclide should appear in the total energy spectrum (M2sum in Fig.~\ref{fig:SpectrumCCVRall}) as a peak at the Q-value of the decay because the total energy ($\alpha$ + nuclear recoil) is  collected by the two facing crystals. 
At the same time, if the surface contamination is deep enough, the $\alpha$ particle can be absorbed by the crystal, giving rise to the same signature of a bulk event.
\subsection{U/Th surface contaminations}
\label{UThsurf}
\begin{figure}
\begin{center}		
\includegraphics[width=0.45\textwidth]{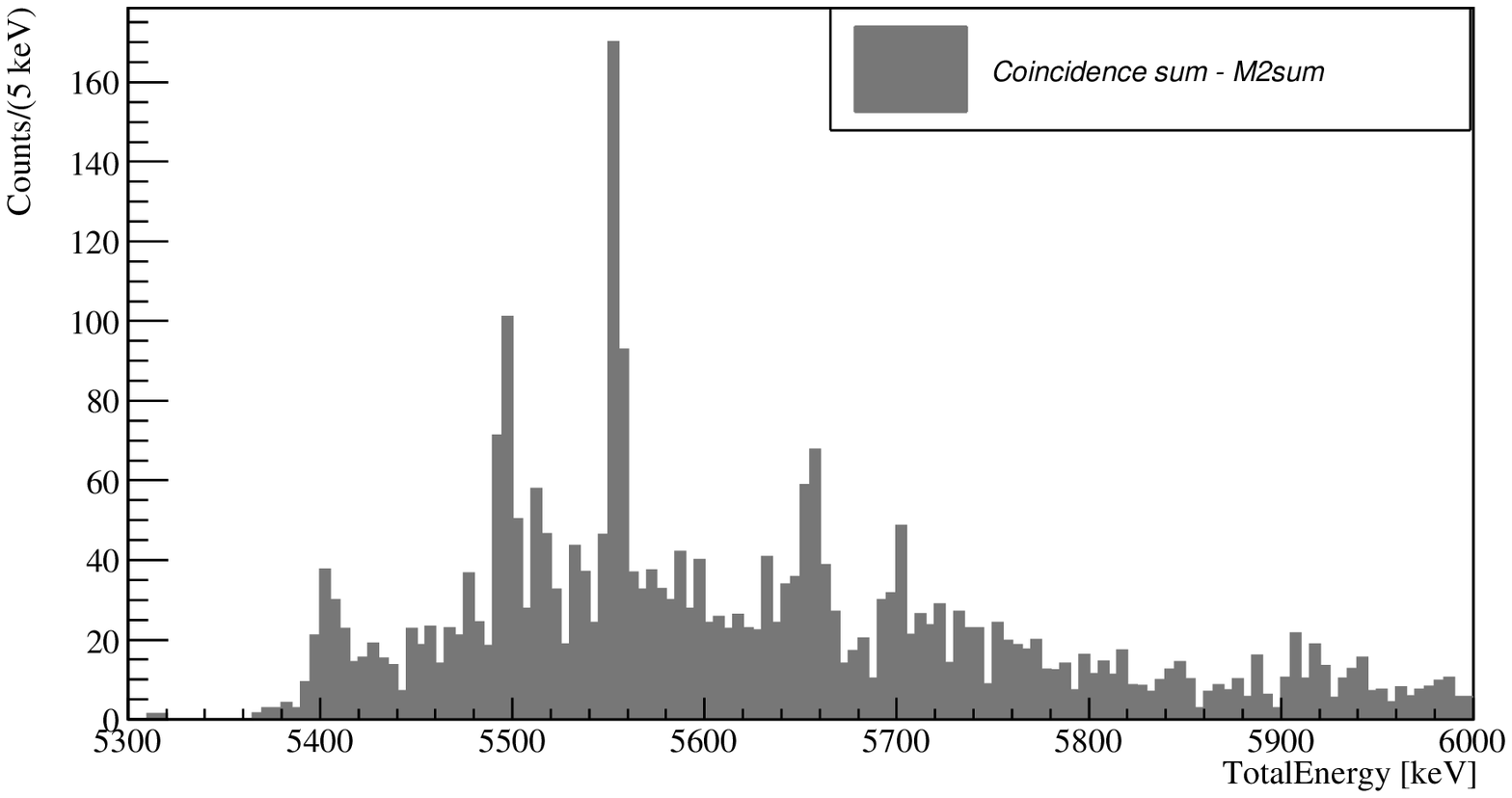}
\includegraphics[width=0.45\textwidth]{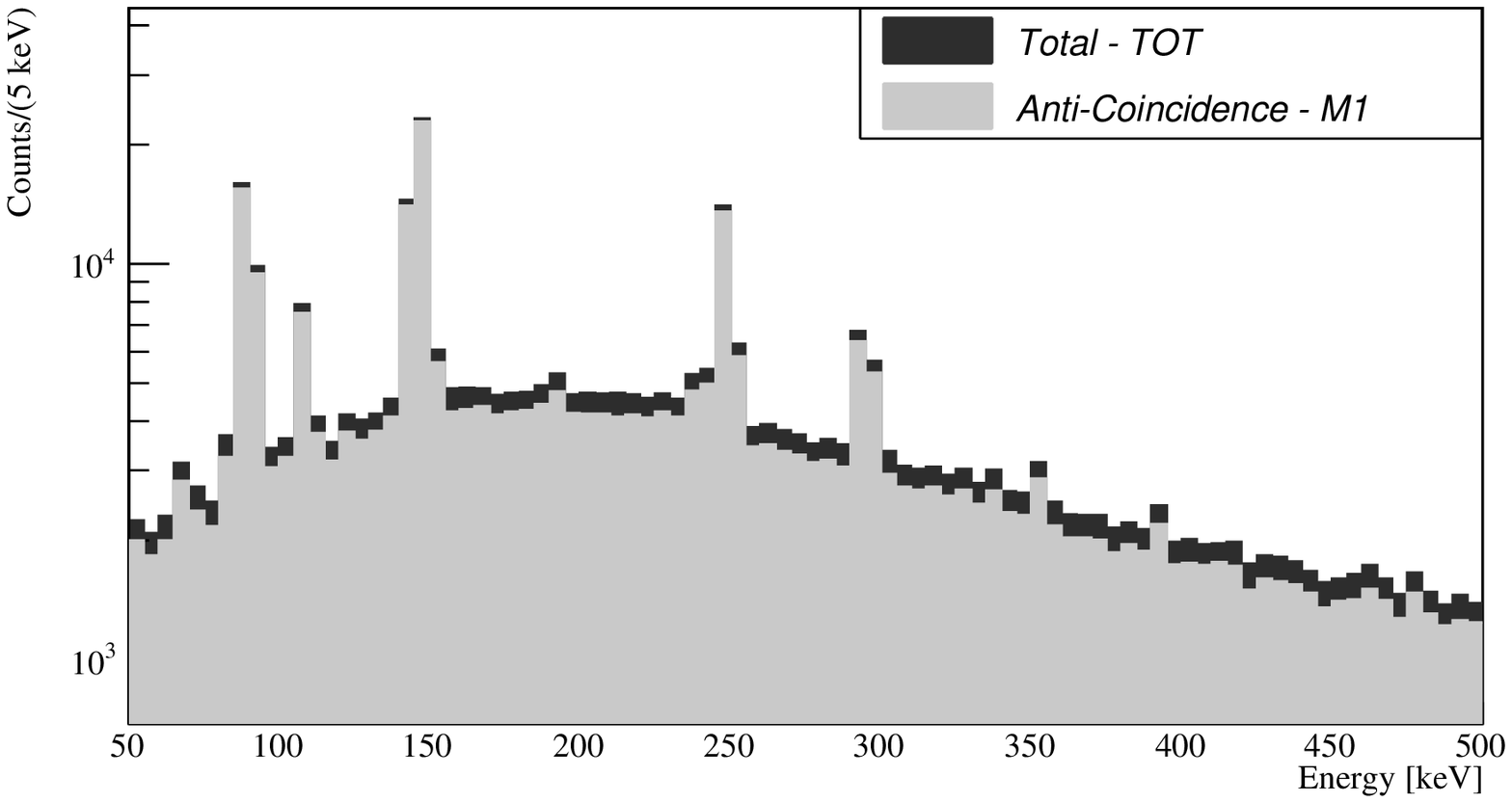}
\end{center}
\caption{Top: M2sum spectrum in the region (5300, 6000) keV. Peaks originated by the coincidence of the 5407.5 keV $\alpha$ line  and a low energy $\gamma$ line from Te metastable isotopes are clearly visible. Bottom: total and anti-coincidence spectra in the region 0-500 keV: low energy lines from Te metastable isotopes are clearly visible.}  
\label{fig:M2sumrandom}
\end{figure}

The M2sum spectrum in Fig.~\ref{fig:SpectrumCCVRall} shows a huge background for energies above the polonium $\alpha$ line due to random coincidences between $^{210}$Po and low energy events. At a closer inspection (see Fig.~\ref{fig:M2sumrandom}),  the peaks originated by the sum of a 5407.5 keV $\alpha$ line and a low energy $\gamma$ line arising from Te metastable isotopes are clearly visible. See also Table~\ref{tab:LowEn} for details.
\begin{table}[htdp]
\begin{center}
  \begin{tabular}{|c|c|c|}
    \hline
    Te isotope & Energy & Energy + 5407.5 \\
    & [keV] & [keV] \\
    \hline
    \hline
    $^{127}$Te & 88.3  & 5495.8\\
    $^{129}$Te & 105.5 & 5513 \\
    $^{125}$Te & 144.8 & 5552.3 \\
    $^{123}$Te & 247.5 & 5655 \\
    $^{121}$Te & 294 & 5701.5 \\
  \hline
  \end{tabular}
  \caption{Energy of the gamma lines from Tellurium metastable isotopes and corresponding sum energy when in coincidence with a $^{210}$Po event.}
  \label{tab:LowEn}
\end{center}
\end{table}

\begin{figure}	
   \centering	
   \includegraphics[width=9cm]{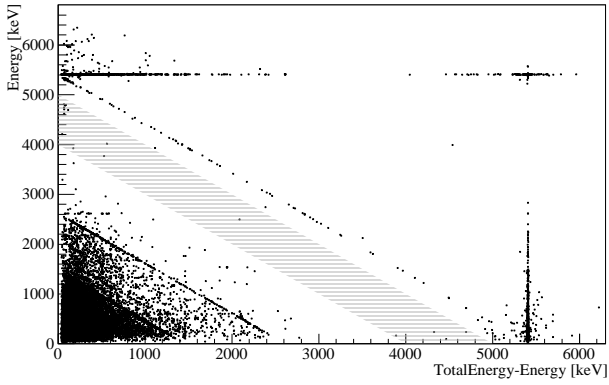}
   \caption{Scatter plot of M2 events. The shadowed region contains events with total energy between 4 and 5 MeV, used for surface analysis.} 
   \label{fig:ScatterPlot}
\end{figure}

Because of the presence of these coincidences, only nuclides with a Q-value lower than 5407.5 keV (see Table~\ref{tab:SurfaceCounts}) are used for surface contamination analysis. $^{218}$Po and $^{212}$Bi are therefore discared. The scatter plot of M2 events is shown in Fig.~\ref{fig:ScatterPlot}. The shadowed region contains the events with a total energy within 4 and 5 MeV. 

For each nuclide listed in Table~\ref{tab:SurfaceCounts}, the number of counts is computed from the M2sum energy spectrum in an energy window of $\pm$ 6$\sigma$ around the Q-value and divided by $\varepsilon^2$ (where $\varepsilon$ is the efficiency of the event-based cuts of  Table~\ref{tab:Effs}), since two coincident events have independent probabilities of passing the cuts. 
In Table~\ref{tab:SurfaceCounts} are shown the corresponding upper limits at 90\% C.L., computed using the Bayesian approach with 0 expected background counts  and a flat prior for the signal~\cite{PDG}.
%

\begin{table}[htdp]
\begin{center}
  \begin{tabular}{|c|c|c|c|c|}
   \hline
    Chain & Nuclide &Half-life  & Energy range & $N_u$ \\
	       &   &[y]        & [keV]        &  90\% C.L.\\  
    \hline
    \hline
           $^{238}$U  & $^{238}$U  &4.47E+09 &4257 - 4283 & 4.6\\ 
                            & $^{234}$U  & 2.45E+05 &   4845 - 4871& 4.6\\ 
		        &  $^{230}$Th &  7.54E+04 &  4757 - 4783 & 4.6\\  
    			&  $^{226}$Ra & 1599 & 4857 - 4883 & 10.5\\

    \hline
    $^{232}$Th &$^{232}$Th & 1.4E+10&4069 - 4095 & 2.3\\
     \hline
    
      \end{tabular}
  \caption{Upper limits at 90\% C.L.  on the number of counts ascribed to several nuclides from uranium and thorium decay chain from the M2sum spectrum. For each nuclide, the half-life of the $\alpha$ decay is also shown.}
  \label{tab:SurfaceCounts}
\end{center}
\end{table}

\begin{figure}	
\centering	
\includegraphics[width=8.5cm]{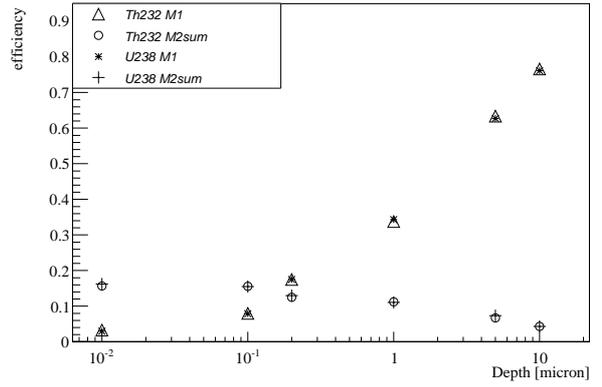}
\caption{For each penetration length, the containment efficiency of surface events is  computed for both M1 and M2sum Monte Carlo spectra.} 
\label{fig:Eff}
\end{figure}

Monte Carlo spectra for $^{238}$U and $^{232}$Th contamination on the crystal surface are generated with exponential profile and various penetration lengths. For each contamination depth, the containment efficiency is calculated in a $\pm$ 6$\sigma$ interval around the Q-value of each nuclide both for the M2sum spectrum and for the M1 spectrum. Results are shown in Fig.~\ref{fig:Eff}. For penetration length of 0.01 and 0.1$\mu$m the higher containment efficiency comes from the M2sum spectrum, whereas for depths of 1, 5 and 10 $\mu$m this arises from the M1 spectrum.  The efficiency containment value for 0.2 $\mu$m depth is very similar both for M1 and M2sum spectrum. 

The drop in M1 containment efficiency for very thin layers of contaminations is due to the fact that, of the about 50\% of decays where the alpha is emitted toward the bulk of the crystals and the recoil is emitted towards outside, in most of the cases the recoil escapes undetected, thus the full Q-value of the alpha decay is not recorded.

The upper limits at 90\% C.L. for the surface activity of each nuclide are evaluated using Eq.~\ref{eq:SurfActivity} where $N_u$ is the 90\% C.L. upper limit on the number of observed events from the M2sum or the M1 spectrum (see Table~\ref{tab:SurfaceCounts} for the M2sum counts and Table~\ref{tab:UThbulkNoC} for the M1 counts) and $\varepsilon_{MC}$ is the Monte Carlo average efficiency defined in eq.~\ref{eq:effMC} for the corresponding spectrum (M2sum or M1) and for the given signature. The confidence intervals for surface activity contaminations are calculated for both the M2sum and M1 spectra, normalizing each signature with the corresponding Monte Carlo efficiency.  The signature giving the most stringent result is taken into account.
 
For surface contaminations of 0.01, 0.1 and 0.2 $\mu$m depths, the most stringent limits come from the M2sum spectrum. For the remaining depths  (1, 5 and 10 $\mu$m) the surface activity reduces practically to a bulk activity and the M1 signature produces the most stringent limits (see Table~\ref{tab:UThbulkNoC} for details).

For the $^{238}$U chain, the surface contamination for 2 peaks is evaluated:
\begin{itemize}
\item $^{238}$U, the chain parent;
\item $^{226}$Ra, the most active line both in M1 and M2sum spectra.
\end{itemize}
The contribution from $^{210}$Pb is treated separately (see Sec.~\ref{sec:210Pb}).

For the $^{232}$Th chain there is only one useable line for this analysis, that is the one from $^{232}$Th. This means that there is no way of testing the portion of the chain below $^{220}$Rn and take into account a possible non-equilibrium of the chain, as  done for the bulk contamination.
 
The results for surface contaminations are shown in Table~\ref{tab:CCVRSurf}. As explained in Sec.~\ref{subsec:UThbulk}, the above confidence intervals are calculated under the hypothesis that the observed counts for each nuclide are entirely due to a surface contamination of that nuclide in the corresponding depth. Again, this is a very conservative hypothesis.

\begin{table}[htdp]
\begin{center}
\small\addtolength{\tabcolsep}{-1pt}
  \begin{tabular}{|c|c|c|}
    \hline
    Depth & Nuclide  & Upper limit 90\% C.L. \\
                 &              & [Bq/cm$^2$]  \\ 
    \hline
    \hline
    $0.01\mu$m  &   $^{238}$U   & 3.1E-09\\  
                                  & $^{226}$Ra & 6.3E-09  \\
                              & $^{232}$Th & 1.6E-09       \\   
       \hline \hline
    $0.2\mu$m &   $^{238}$U   & 3.8E-09\\  
                                  & $^{226}$Ra & 7.6E-09  \\
                              & $^{232}$Th & 2.0E-09       \\   
       \hline \hline
    $1\mu$m &   $^{238}$U   & 3.7E-09\\  
                                  & $^{226}$Ra & 8.9E-09  \\
                              & $^{232}$Th & 1.9E-09       \\   
  \hline \hline
    $5\mu$m &   $^{238}$U   & 2.0E-09\\  
                                  & $^{226}$Ra & 5.4E-09  \\
                              & $^{232}$Th & 1.0E-09       \\   
  \hline \hline
    $10\mu$m   & $^{238}$U    & 1.7E-09\\  
                          & $^{226}$Ra  & 4.4E-09  \\
                          & $^{232}$Th   & 8.3E-10      \\   
    \hline   
  \end{tabular}

  \caption{Upper limits at 90\% C.L. for surface contamination, for different penetration length values. See text for details on the calculation of confidence intervals.}\label{tab:CCVRSurf}
\end{center}
\end{table}
\section{Extrapolation to CUORE background}
\label{sec:extrCuore}
In order to evaluate the contribution to the CUORE background arising from crystal contamination, a simulation both for bulk and surface contamination reproducing the CUORE geometry and studying the contribution to the double beta decay (DBD) energy region is performed. 

Being interested in a conservative upper limit to the CUORE background, in this extrapolation it has been assumed that CUORE crystals will have the same activity as crystals tested in CCVRs. However, when CUORE will start the data taking, most of the $^{210}$Po (and other short lived states) will have decayed and the efficiency for pile-up rejection will be increased.

\subsection{Background from bulk contamination}
The CUORE geometrical efficiency for a uniform bulk contamination of crystals in $^{210}$Pb, $^{238}$U and $^{232}$Th is estimated via Monte Carlo, and the corresponding CUORE background at the DBD energy region (Q-value $\pm$ 30 keV) is calculated using the formula:
\begin{equation}
{\rm bkg}_{{\rm CUORE}} = \frac{A_{{\rm CCVR-bulk}} \; \varepsilon^{{\rm CUORE-bulk}}_{MC}} {\Delta E}
\end{equation}
where $A_{{\rm CCVR-bulk}}$ are the values of  $^{210}$Pb, $^{238}$U and $^{232}$Th activities from Tables~\ref{tab:UThbulkEq} and ~\ref{tab:CCVR1Pb210}, $\varepsilon^{{\rm CUORE-bulk}}_{MC}$ is evaluated from Monte Carlo simulation and $\Delta$E = 60 keV. 

%
\begin{table}[htdp]
\begin{center}
\small\addtolength{\tabcolsep}{-1pt}
  \begin{tabular}{|c|c|c|}
 \hline
 chain &Nuclide & Upper Limit 90\% C.L. \\
	&        &[counts/keV/kg/y] \\
    \hline
    \hline
       
       & $^{210}$Pb  &   2.6E-05 \\
       \hline
       $^{238}$U & $^{238}$U  & 8.1E-07 \\
                        & $^{234}$U     & 1.4E-06 \\
                        & $^{230}$Th   & 1.7E-06\\
                        & $^{226}$Ra  & 2.1E-06\\
                        & $^{218}$Po   & 5.1E-07\\
    \hline
    $^{232}$Th & $^{232}$Th & 1.7E-05 \\
                          & $^{212}$Bi  & 1.1E-04 \\ 
     \hline
  \end{tabular} 
 \caption{Extrapolation to CUORE background from CCVRs bulk contamination limits from Tabs.~\ref{tab:UThbulkEq} and~\ref{tab:CCVR1Pb210}.} 
  \label{tab:CUOREbulk}
\end{center}
\end{table}

The results are shown in Table~\ref{tab:CUOREbulk}. \\
In the most conservative approach, considering the most active line ($^{212}$Bi), the following upper limit to the CUORE background at the DBD energy due to bulk contamination of crystals is set: 1.1 $\cdot$ 10$^{-4}$ counts/keV/kg/y.

\subsection{Background from surface contamination}
In a similar way, the CUORE geometrical efficiency for a surface contamination for several depths is estimated. The corresponding CUORE background is extrapolated using the formula:
\begin{equation}
{\rm bkg}_{{\rm CUORE}}= \frac{A_{{\rm CCVR-surf}} \; \, \varepsilon^{{\rm CUORE-surf}}_{MC} \; S} {\Delta E \; M_{{\rm CUORE}}}
\end{equation}
where $A_{{\rm CCVR-surf}}$ are the surface contamination values from Tables~\ref{tab:CCVR1Pb210} and~\ref{tab:CCVRSurf}, $\varepsilon^{{\rm CUORE-surf}}_{MC}$ is estimated via Monte Carlo, $S$ is the surface of the CUORE crystals, $\Delta$E = 60 keV and $M_{{\rm CUORE}}$ = 0.75 $\cdot$ 988 kg.

\begin{table}[htdp]
\begin{center}
\small\addtolength{\tabcolsep}{-1pt}
 \begin{tabular}{|c|c|c|c|}
    \hline
    Depth & Nuclide  & Upper limit 90\% C.L.\\
                 &              & [counts/keV/kg/y]   \\ 
    \hline
    \hline
    $0.01\mu$m  &  $^{210}$Pb   & 1.6E-03\\  
    	&   $^{238}$U   & 8.9E-05\\  
                               & $^{226}$Ra & 1.8E-04  \\
                              & $^{232}$Th & 1.8E-05       \\   
       \hline \hline
    $0.2\mu$m &  $^{210}$Pb   & 2.3E-04\\  
    &   $^{238}$U   & 3.9E-04\\  
                               & $^{226}$Ra & 7.7E-04  \\
                              & $^{232}$Th & 1.5E-04       \\     
       \hline \hline
    $1\mu$m & $^{210}$Pb   & 3.8E-04\\ 
     &     $^{238}$U   & 1.2E-03\\  
                               & $^{226}$Ra & 2.9E-03  \\
                              & $^{232}$Th & 4.4E-04       \\     

  \hline \hline
    $5\mu$m &     $^{210}$Pb   & 4.3E-04\\   
     & $^{238}$U   & 1.3E-03\\  
                               & $^{226}$Ra & 3.4E-03  \\
                              & $^{232}$Th & 4.7E-04       \\     
      \hline \hline
    $10\mu$m  & $^{210}$Pb   & 2.9E-04\\  
    & $^{238}$U   & 7.1E-04\\  
                               & $^{226}$Ra & 1.9E-03  \\
                              & $^{232}$Th & 3.0E-04       \\     
        \hline   
  \end{tabular}
  \caption{Extrapolation to CUORE background from CCVRs surface contamination limits from Tables~\ref{tab:CCVR1Pb210} and ~\ref{tab:CCVRSurf}.}\label{tab:CUOREsurf}
\end{center}
\end{table}

\begin{figure}	
   \centering	
   \includegraphics[width=8cm]{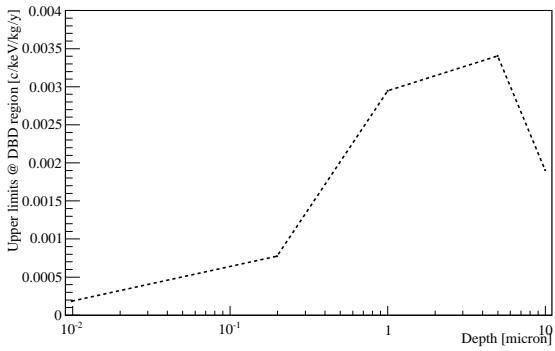}
   \caption{For each penetration depth, the upper limits at  90\% C.L. for the count rate at the DBD energy region  are plotted. The values are computed on $^{226}$Ra line, and are due only to surface crystals contaminations.} \label{fig:Bkgdepth}
\end{figure}

The results are shown in Table~\ref{tab:CUOREsurf}. Based on these results, in the most conservative approach, a crystal surface contamination in CUORE within the limits estimated from CCVRs data, would yield a rate in the DBD energy region lower than 5.5\,$\cdot$\,10$^{-3}$ counts/keV/kg/y. This value corresponds to the worst case: a contamination in $^{238}$U and  $^{232}$Th at 5$\mu$m and a contamination in $^{210}$Pb at 0.01$\mu$m.
This result is strictly dependent on the density profile assumed for the contaminants distribution, as can be seen in Fig.~\ref{fig:Bkgdepth}, where the upper limit at the 90\% C.L.  for the rate in the DBD energy region (computed on $^{226}$Ra line) is plotted in function of contamination depth.

\section*{Conclusions}
The CUORE Crystal Validation Runs (CCVRs) have been carried out since the end of 2008 at the Gran Sasso National Laboratories, in order to test the performances and the \ radiopurity of the TeO$_2$ crystals produced at SICCAS  (Shanghai Institute of Ceramics, Chinese Academy of Sciences) for the CUORE experiment. 

In this work the results of the first 5 validation runs, are presented. The items under analysis are the background rate, the bulk and surface contamination from several nuclides and the extrapolation to the CUORE background. 

The CCVRs background rate shows a reduction with respect to the one measured in CUORICINO in all the energy regions considered. The bulk activity of $^{210}$Po is measured to be within the limit specified in the contract with the crystals producer. No indication of a bulk contamination from uranium and thorium decay chains, as well as from $^{210}$Pb (out of equilibrium), is found. The upper limits are calculated to be within the contract specification. No indication of a surface contamination from uranium and thorium decay chains, as well as from $^{210}$Pb (out of equilibrium), is found. Upper limits at 90\% C.L. are calculated for surface contamination from several nuclides and for different contamination depths.

An extrapolation to CUORE background from both bulk and surface contaminations is performed, in the most conservative assumption that the CCVRs observed background is entirely due to the bulk and the surface contamination respectively. Considering only the contribution from bulk and surface crystals contaminations, the following upper limits to the CUORE background index in the energy region around the Q-value of the neutrinoless double beta decay of $^{130}$Te are calculated: 1.1\,$\cdot$\,10$^{-4}$ counts/keV/kg/y and 5.5\,$\cdot$\,10$^{-3}$ counts/keV/kg/y respectively. 

\begin{thebibliography}{99}
\bibitem{CUORE} R.~Ardito et al., arXiv:hep-ex/0501010.
\bibitem{CUORE04}C.~Arnaboldi et al., Nucl. Instr. Meth. A 518 (2004) 775.
\bibitem{Arna10} C.~Arnaboldi et al., Journal of Crystal Growth 312 (2010) 2999.
\bibitem{NTD} E.E. Haller et al., edited by R. D. Larrabee (Plenum Press, New York, 1984) 21.
\bibitem{Qino08} C.~Arnaboldi et al., Phys. Rev. C 78 (2008) 035502.
\bibitem{HallC} S.~Pirro, Nucl. Instr. Meth. A 559 (2006) 672.
\bibitem{Qino10} E.~Andreotti et al., Astr. Phys. 34 (2011) 822.
\bibitem{Gatti} E.~Gatti and P.F.~Manfredi, Rivista del Nuovo Cimento 9 (1986) 1.
\bibitem{ArtRadio} C.~Bucci et al., Eur. Phys. J A 41 (2009) 155.
\bibitem{ArtChambery} M.~Pavan et al., Eur. Phys. J A 36 (2008) 159.
\bibitem{TTT} CUORE Coll. "Three Tower Test: copper surface treatments for low background DBD experiments with TeO$_2$ bolometers." Paper in preparation.
\bibitem{PDG} K.~Nakamura et al. (Particle Data Group), J. Phys. G 37, 075021 (2010)
\end{thebibliography}
\end{document}